# Malware Detection Using Frequency Domain-Based Image Visualization and Deep Learning

Tajuddin Manhar Mohammed
Mayachitra, Inc.
Santa Barbara, California
mohammed@mayachitra.com

Lakshmanan Nataraj
Mayachitra, Inc.
Santa Barbara, California
nataraj@mayachitra.com

Satish Chikkagoudar
U.S. Naval Research Laboratory
Washington, D.C.
satish.chikkagoudar@nrl.navy.mil

Shivkumar Chandrasekaran
Mayachitra, Inc., and
ECE Department, UC Santa Barbara
Santa Barbara, California
shiv@ece.ucsb.edu

B.S. Manjunath
Mayachitra, Inc., and
ECE Department, UC Santa Barbara
Santa Barbara, California
manj@ece.ucsb.edu

## Abstract

*We propose a novel method to detect and visualize malware through image classification. The executable binaries are represented as grayscale images obtained from the count of N-grams (N=2) of bytes in the Discrete Cosine Transform (DCT) domain and a neural network is trained for malware detection. A shallow neural network is trained for classification, and its accuracy is compared with deep-network architectures such as ResNet that are trained using transfer learning. Neither dis-assembly nor behavioral analysis of malware is required for these methods. Motivated by the visual similarity of these images for different malware families, we compare our deep neural network models with standard image features like GIST descriptors to evaluate the performance. A joint feature measure is proposed to combine different features using error analysis to get an accurate ensemble model for improved classification performance. A new dataset called MaleX[1] which contains around 1 million malware and benign Windows executable samples is created for large-scale malware detection and classification experiments. Experimental results are quite promising with 96% binary classification accuracy on MaleX. The proposed model is also able to generalize well on larger unseen malware samples and the results compare favorably with state-of-the-art static analysis-based malware detection algorithms.*

## 1. Introduction

Malicious applications and software (malware) are one of the major security threats the internet faces today. To detect malware with high precision and less computational power is beneficial as it becomes significantly easier to create strategies to assuage the extent of harm caused by such malware.

Many researchers use hand-crafted features to detect malware (discussed in Section 2), but recent advancements in machine learning suggest that deep neural networks are superior in learning useful features and come up with good malware representations. These features are at best, non-orthogonal and significantly improve the accuracies of classification, compared to human hand-crafted features. In this paper, we approach the above problem of detecting malware with a Convolutional Neural Network (CNN) model. In this paper, we apply CNNs to the problem of malware detection.

Previous research has shown that malware variants belonging to the same family exhibit visual similarity in the byteplot images [1, 2]. These are based on visualization in the spatial domain by converting bytes to pixels. In this paper, we take a different approach and visualize malware in the frequency domain to detect malware. The motivation to visualize malware in frequency domain is because of the "sparse" feature representations of malware in literature [3, 4, 5, 6, 7, 8] that are typically extracted from raw bytes of the binaries or disassembled instructions (n-grams, n-perms). One way to visualize such sparse signals is to compute Discrete Cosine Transform (DCT) of the signal to "de-sparsify" them and get a better visual representation which will be useful for various applications like classification and detection. We present one such method which visualizes frequency count of bi-grams of raw bytes in frequency domain. After obtaining these digital image representations of malware binaries, we employ Convolutional Neural Networks to detect malware. Finally, we also designed a joint feature metric (Appendix. A) that justifies

---

[1] https://github.com/Mayachitra-Inc/MaleX

combining both these features to create an accurate ensemble model for malware detection.

The main contributions in this paper are as follows:

- We propose a novel method to detect malware by visualizing malware binaries in the frequency domain ("bigram-dct" images).

- We created a new dataset called *MaleX* which has 864,669 malware and 179,725 benign Windows executable samples that can be used for large data-driven malware detection and classification experiments.

- Leveraging recent advances in machine learning, we employ Convolutional Neural Networks (CNNs), which are known to achieve superior accuracy, to detect malware.

- Finally, we combine features from byteplot images and "bigram-dct" images using the proposed joint feature metric to get a highly accurate ensemble neural network model for malware detection.

The rest of the paper is organized as follows. In Section 2, we discuss the related work in malware detection and visualization. In Sections 3 and 4, we present our method to visualize malware binaries in frequency domain and detect them using the proposed Convolutional Neural Network (CNN) model. We discuss the evaluation metrics, experimental methods, results and discussion in Section 5 and conclude in Section 6. In Appendix. A, we define an "orthogonality" metric to better understand the combination of different features for improved classification performance.

## 2. Related Work

Typical features extracted from malware can be broadly grouped into either *static* or *dynamic* features. As the name suggests, static features are extracted from the malware without executing it. Dynamic features, on the other hand, are extracted by executing the code, usually in a virtual environment, and then studying their behavioral characteristics such as system calls trace or network behavior. We will review some of the approaches here.

Static analysis can further be classified into static code based analysis and non code based analysis. Static code based analysis techniques study the functioning of an executable by disassembling the executable and then extracting features. The most common static code based analysis method based is control flow graph analysis [9, 10, 11, 12, 13]. After dis-assembly, the control flow of the malware is obtained from the sequence of instructions and graphs are constructed to uniquely characterize the malware. Static non code based techniques are based on a variety of techniques: n-grams [14, 15, 16, 17, 18], n-perms [7, 8, 19], hash based techniques [20, 21, 22], PE file structure [23, 24, 25, 26] or signal similarity based techniques [1, 27, 28, 29, 30]. Among the first two methods, n-grams or n-perms are either computed on the raw bytes of the binaries or on the disassembled instructions. Features are then extracted from these to characterize the malware. However, n-gram-based approaches are less scalable because of the computationally expensive feature matching operation over relatively large dimensionality of the n-gram feature space. Jang et al. [5] proposed feature hashing to reduce the high-dimensional feature space in malware analysis, and implemented feature hashing on n-gram based features.

Among hash based methods, *ssdeep* [31] is a common technique to compute context triggered piece-wise hashes on raw binaries. *Pehash* [32], however, uses the Portable Executable (PE) file structure to compute a similarity hash. There are also techniques that extract discriminative features from the Portable Executable (PE) structure of an executable [33, 34, 35]. Image similarity based methods [1, 27] convert a malware binary to a digital image and apply image processing based techniques to compute similarity features. Among dynamic analysis, the most common method is to execute the malware in a controlled environment and then study its execution behavior. Behavioral profiles or graphs are generated to build models of malware [36, 37]. Some works generate a human readable report of the execution flow and extract features from the reports [38, 39].

Recently, there has been lot of research involving malware detection using deep learning [40, 41], recurrent neural networks (RNNs) [42, 43, 44], convolutional neural networks (CNNs) [45, 46, 47, 48, 49] and hybrid models [50, 51, 52]. The works that use CNNs, typically converts malware binaries to digital images and pass them into a CNN in order to detect malware. However, in this paper, we visualize malware in the frequency domain and then use CNNs for malware detection. The proposed malware detection approach also has all the positives of static analysis methods while addressing some of the limitations of the aforementioned works like high time complexity, low scalability and high total feature selection count.

## 3. Malware visualization in frequency domain

A given executable binary file is read as a 16-bit signed hexadecimal vector and divided into corresponding bi-grams (n-grams of bytes with n=2). As an example, for the byte-stream *0a1bc48a*, the corresponding bi-grams will be *0a1b*, *1bc4* and *c48a*. We then use the bi-gram frequency count to get a sparse image of dimensions 256x256 (where each pixel intensity value corresponds to normalized frequency count of a particular bi-gram). The image dimensions is the primary reason for choosing bi-grams (n=2) as it aids image-based machine learning classification with low time complexity in the training phase. Finally, we compute the full frame Discrete Cosine Transform (DCT) of this image to de-sparsify and get a resulting "bigram-dct" image with distinctive textured patterns. Before normalizing frequency count to get the sparse image, we zero out the first frequency count value corresponding to the bi-gram *0000*, as this value is relatively very large compared to that of other bi-grams and the bi-gram either corresponds to empty space or new line(s) in the code with no useful information. The overview of the proposed feature extraction method is shown in Figure 1.

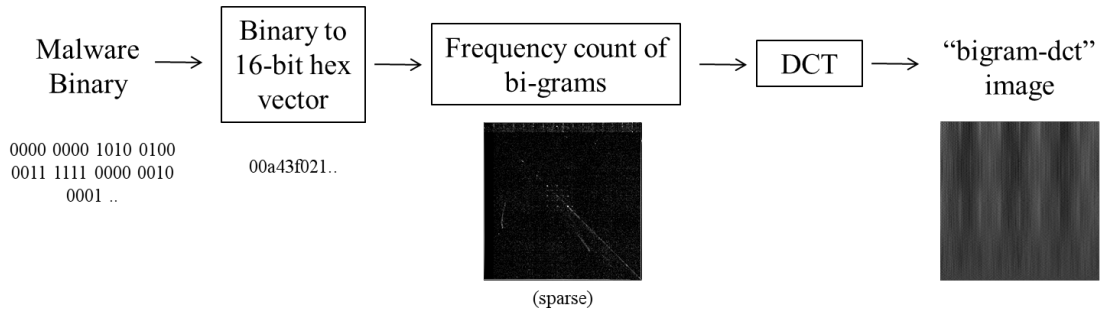

Figure 1: Visualizing malware as a grayscale image in DCT domain.

Even-though, in this work, we restrict our discussion to malware detection rather than malware classification, we see (Figure 2) that these textured patterns are distinct for various malware families and can also be used to classify malware binaries into different malware families.

## 4. Malware Detection using CNN

Deep learning recently gained remarkable success in solving computer vision problems, especially the task of image classification. For our "bigram-dct" images, one can extract global image features like GIST [54], which have been shown to be successful in scene classification and object classification [55] and feed them into standard classifiers such as k-NN (k-Nearest Neighbors) or Random Forests to detect malware binaries through classification. These features are computed by convolution of the filter with an image at different scales and orientations. Thus, high and low frequency repetitive gradient directions of an image can be measured using GIST features. But, to perform classification in an end-to-end manner and to be more discriminative, one can utilize Convolutional Neural Networks (CNNs) to automatically extract relevant features, similar to what Razavian et al. did in [56].

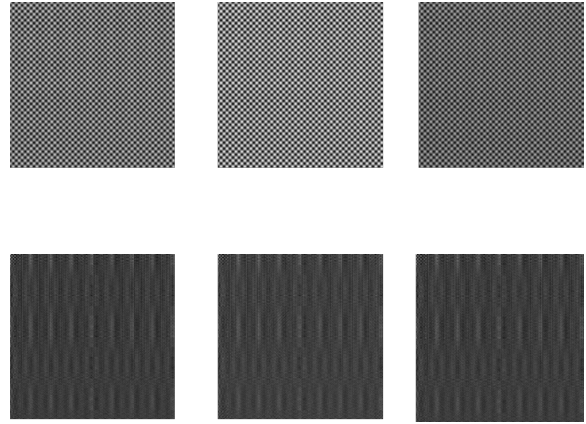

Figure 2: The images in the first row are "bigram-dct" images of three instances of malware belonging to the family Lolyda.AT and those in the second row belong to the family Adialer.C (malware samples from the MalImg [53] dataset).

### 4.1. Network Architectures

We use a shallow network to have more control over the design and parameters of the neural network. We also used pre-trained ImageNet [57] deep network models (ResNet-18/ResNet-50) [58] and fine-tuned them to observe the effect of depth of networks on the feature learning paradigm. Both the network architectures are briefly described below.

**4.1.1. Proposed "3C2D" model** Eventhough, Simonyan et al. [59] showed that classification accuracies can be improved by going deep in the network, we used a shallow CNN model similar to

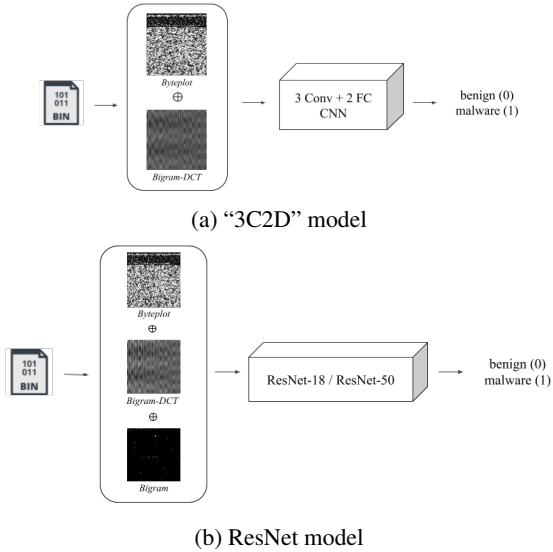

(a) "3C2D" model

(b) ResNet model

Figure 3: Pipeline for the end-to-end deep learning-based malware detection.

LeNet-5 model [60] to have control over the model in terms of number of parameters and time complexity. Our proposed CNN architecture consists of:

1. Input layer of 256x256 pixels
2. Convolutional layer (32 filter maps of size 3x3)
3. Max-pooling layer (size 2x2)
4. Convolutional layer (64 filter maps of size 3x3)
5. Max-pooling layer (size 2x2)
6. Convolutional layer (128 filter maps of size 3x3)
7. Max-pooling layer (size 2x2)
8. Densely-connected layer with Dropout (512 neurons)
9. Densely-connected layer with Dropout (256 neurons)
10. Output "sigmoid" layer

A potential threat for any deep neural network model is overfitting. Srivastava et al. [61] invented 'dropout' where the units in the layers are randomly dropped to prevent overfitting. We used two such dropout layers (with probability = 0.5) after the two fully-connected layers to prevent overfitting. In testing phase, these dropout layers will be removed.

This model is used for both the binary classification (end-to-end pipeline outlined in Figure 3a) and the performance comparison (with the standard classifiers) experiments, as will be discussed in the following section.

**4.1.2. ResNet model** Almost all the transfer learning-based approaches in the domain of malware detection and classification [62, 63, 64, 65, 66] use deep network architectures that are known to perform well for image classification task. We employed similar principles and used well-known ResNet variants – ResNet-18 and ResNet-50 for the purpose of malware classification.

As the model expects a 3-channel input, we use sparse and dct-based representations of bigram images in addition to the well-known byteplot [53] image representation of malware binaries (end-to-end pipeline outlined in Figure 3b). More details are discussed in the following section.

This model is used for comparing classification performance to that of the proposed shallow CNN model and standard classifiers.

## 5. Experiments and Results

We conducted our experiments on two different datasets: (1) *smalldata* to parameterize the deep network model, and (2) *MaleX* to show the efficacy of our model. These datasets are described briefly below. The experiments include: (1) Binary classification using the proposed CNN model, and (2) Binary classification of the GIST features using standard classifiers for comparative analysis.

### 5.1. Datasets

We performed our experiments on the following curated datasets.

- *smalldata*: This dataset consists of 16,358 Windows executable samples: 9,130 of which are malware samples from the MalImg dataset ([53]) and the other 7,228 files are benign (clean) samples collected from Windows system files. We also verified the binary labels using VirusTotal [67] service which has an aggregate of many anti-virus products and online scan engines to check for malware.

- *MaleX*: We created this large dataset where benign and malware binaries were drawn from various sources ([68, 69, 70]). Our final dataset after VirusTotal [67] filtering (any sample with VirusTotal detection ratio ≤10% is considered as benign, and ≥90% is considered as malware), and file-type filtering (MIME type: "application/x-dosexec") contains 1,044,394 Windows executable binaries with 864,669 labelled as malware and 179,725 as benign. We believe that this dataset has reasonable number of samples and is sufficient to test data-driven machine learning classification methods and also to measure the performance of the designed models in terms of scalability and adaptability. The only dataset comparable to *MaleX* is *EMBER* [71], but the dataset does not have access to the executable binaries that is needed for our experiments. The binaries of the *MaleX* dataset is available to public upon request.

### 5.2. General Settings

Out of the standard classifiers, we chose k-Nearest Neighbors (k-NN) and Random Forest as they had better classification accuracies than other methods. These algorithms were implemented in scikit-learn [72]. The hyper-parameters were optimized using grid search on *smalldata*. The binary classification accuracies reported in the following sections are the 10-fold cross-validation accuracies.

The proposed CNN models (4.1) were implemented in Keras with TensorFlow backend. We trained different models using 179,725 benign samples and different subsets of 179,725 malware samples. These models were trained with the data partitioned as follows: 70% of images in each class were used for training, 20% for validation, and the remaining 10% for testing. No additional data augmentation method was used. The binary classification accuracies on the test set averaged for different trained models (similar to k-fold cross-validation) are reported in the following sections.

### 5.3. Results

**5.3.1. Binary classification on "bigram-dct" images** Table. 1 shows the classification accuracies on both the datasets. The k-NN and Random Forest models correspond to classification using the 320-dimensional GIST features extracted from the "bigram-dct" images. The CNN model corresponds to our proposed "3C2D" network architecture (4.1.1) with single-channel "bigram-dct" images as inputs.

| Model | *smalldata* | *MaleX* |
|---|---|---|
| k-NN | 0.99083 | 0.83817 |
| Random Forest | 0.98839 | 0.83308 |
| Proposed CNN ("3C2D") | **1.00000** | **0.89491** |

Table 1: Classification accuracies on "bigram-dct" images.

**5.3.2. Binary classification on byteplot images** In our previous work, Nataraj et al. [53] proposed a method for classifying malware represented as byteplot images. Table. 2 shows the classification accuracies using the same models on these byteplot images. The byteplot images were re-sized to 256x256 to maintain consistency among different models and experiments.

Moreover, we define a joint feature metric using error analysis to determine how these two different features (one from "bigram-dct" images and the other from byteplot images) perform together in detecting malware. More details on the metric can be found in Appendix A. The $JFS$ values were 0.87 and 0.81 on *smalldata* and *MaleX* datasets, respectively.

| Model | *smalldata* | *MaleX* |
|---|---|---|
| k-NN | 0.99511 | 0.83405 |
| Random Forest | 0.99022 | 0.82793 |
| Proposed CNN ("3C2D") | **0.99633** | **0.83603** |

Table 2: Classification accuracies on byteplot images.

**5.3.3. Binary classification on "bigram-dct" AND byteplot images** We use both the features and define our ensemble models as follows – the k-NN and Random Forest models correspond to classification using the 640-dimensional GIST features obtained by concatenating the GIST features extracted from both the "bigram-dct" and byteplot images. Similarly, the CNN model has the same architecture as before, but now, the input layer has two channels (as shown in Figure 3a) i.e., the input layer is of dimension 256x256x2. Table. 3 shows the classification accuracies on these ensemble models.

| Model | *smalldata* | *MaleX* |
|---|---|---|
| k-NN | 0.99633 | 0.85447 |
| Random Forest | 0.99267 | 0.84384 |
| Proposed CNN ("3C2D") | **1.00000** | **0.96154** |

Table 3: Classification accuracies on ensemble model (as shown in Figure 3a) including both "bigram-dct" and byteplot images.

**5.3.4. Generalizability of proposed CNN model** To validate the generalization capability of the 2-channel CNN model, we re-trained the models using the above-mentioned train and validation subsets (new training set) and test subset (new validation set), and tested on rest of the left-out *MaleX* dataset which has only malicious samples. Here, we also show the ROC (receiver operating characteristic) curves and report the corresponding AUCs (area under the curve) to show more details about the classifier performance, as shown in Table. 4. Since the test set has only one class of samples, we show ROC for the entire dataset (train + validation + test) and compute accuracy of the test set from the different AUCs recorded.

| Subset | # Malicious Samples | # Benign Samples | AUC |
|---|---|---|---|
| Train | 161,752 | 161,753 | 0.9999252 |
| Validation | 17,973 | 17,972 | 0.9155202 |
| Test | 648,946 | NA | NA |

Table 4: Data distribution of *MaleX* dataset showing classification performance and generalization capabilities of proposed CNN ("3C2D") model.

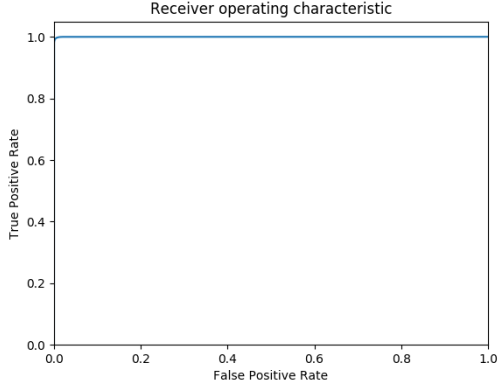

(a) Training set AUC: 0.9999252

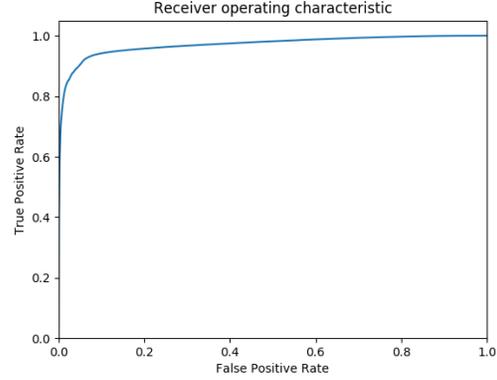

Figure 5: ROC curve for shallow CNN classification model on training, validation and test samples of *MaleX* (AUC: 0.9706473).

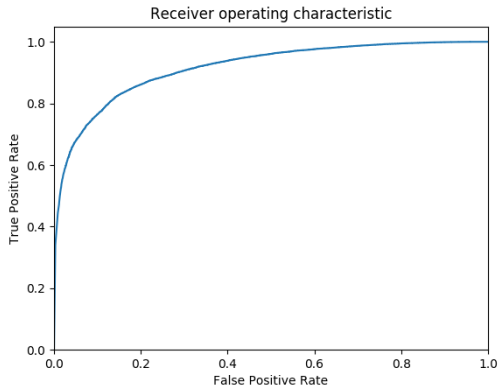

(b) Validation set AUC: 0.9155202

Figure 4: ROC curves for shallow CNN classification model on training and validation samples of *MaleX*.

ROC curve and AUC for train and validation set is shown in Figure 4. ROC and AUC of the model tested on all of *MaleX* dataset is shown in Figure 5. We use a threshold of 0.5 to classify any sample into malicious or benign - a value greater than 0.5 corresponds to malicious samples and less than 0.5 as benign.

#### 5.3.5. Transfer learning-based classification and overall performance

To observe the the effect of neural network model depth on the classification performance, we use pre-trained ImageNet weights, and fine-tune ResNet models (as shown in Figure 3b) to effectively use transfer learning-based classification technique that is well known in literature.

Table. 5 shows the classification accuracies on three different subsets of *MaleX* (as discussed in Sec. 5.3.4) using various classifiers including the proposed CNN and the transfer learning-based ResNet models. Note that the accuracies are not reported for k-NN and Random Forest models on train subset as they are supervised data-driven algorithms that will give accuracies close to (if not exactly) 100%, which will not be useful for performance comparisons.

| Model | Train | Validation | Test |
|---|---|---|---|
| k-NN (byteplot GIST) | NA | 0.7896 | 0.5173 |
| Random Forest (byteplot GIST) | NA | 0.7762 | 0.5039 |
| k-NN ("bigram-dct" GIST) | NA | 0.7905 | 0.5224 |
| Random Forest ("bigram-dct" GIST) | NA | 0.7880 | 0.5204 |
| Proposed CNN | 0.9456 | 0.8451 | 0.8360 |
| ResNet-18 | 0.9566 | 0.8843 | 0.8252 |
| ResNet-50 | **0.9880** | **0.8956** | **0.8620** |

Table 5: Classification performance on different subsets of *MaleX* with different classifiers including the proposed CNN ("3C2D") and the transfer learning-based ResNet models.

### 5.4. Discussion

As shown in Table. 1 and Table. 2, the proposed approach using 'bigram-dct' images and CNNs outperforms the features obtained from byteplot images. Please note that we also used the spatial bigram features (results not reported) which performed poorly in our classification experiments and that was the motivation for using 'bigram-dct' images in the first place. The high joint feature metric scores mentioned in Section 5.3.2 suggest that the features obtained from "bigram-dct" images can be combined with the ones from byteplot images with high confidence to give a better representation of the malware in its entirety.

We also ensure that there is neither variance-bias nor overfitting while training the model. We see a significant improvement in the classification accuracies on the ensemble model that uses both the "bigram-dct" and byteplot images. This also justifies the use of our joint feature metric to combine these features. We also see that our proposed CNN model outperforms standard classifier models that uses global GIST features in all the cases.

Section 5.3.4 discusses behavior of the trained model towards unseen larger malware subset. It is evident from the AUC scores, that the model is learning and is able to achieve good performance on validation set. On the test set which has only malicious samples, our model achieves accuracy of $83.60\%$. This is a "good" score that signifies the generalization capability of the model, given that the trained model has a shallow network architecture and has learned from relatively less number of samples (training set has 161,752 malicious samples compared to that of 648,946 malicious samples in the test set).

In Section 5.3.5, we use transfer learning-based approach to train deeper networks and see significant improvements in the overall classification performance, as shown in Table. 5. This is generally the case in computer vision literature that addresses the task of image classification. This also signifies the fact that the deeper networks have advantage of having multiple layers that can learn features at various levels of abstraction, making them much better at generalizing to new data. The table also summarizes all the nitty-gritty findings of the experiments and gives a concise representation of the overall malware detection performance using the DCT image-based representations derived from the malware binaries.

## 6. Conclusion

In this work, we propose a malware detection mechanism using byteplot images and grayscale images from the count of bi-grams of bytes in DCT domain, and a deep learning technique that involves Convolutional Neural Networks (CNNs). Our results confirm that visual textures and patterns in these images can be used for accurate malware detection. We evaluated our model on equi-proportionate subsets of large dataset we created called *MaleX* consisting of Windows executable samples, obtaining a mean binary classification accuracy of $96\%$. We also demonstrate the generalization capability of this model by showing the classification performance on the bigger left-out *MaleX* samples.

Moreover, we demonstrated the use of deep learning that includes transfer learning-based approach in malware detection tasks and showed that the features learned by the deep CNN models (ResNet) performed significantly better than the hand-crafted GIST features proposed in the literature, and also slightly better than the features learned by the proposed shallow CNN model. Finally, we also designed a metric to measure the performance of the ensemble model which uses combination of different features that was used to include both the byteplot image and the "bigram-dct" image as inputs to our designed neural network model.

## Acknowledgement

This work has been supported by the ONR contract #N68335-17-C-0048. The views expressed in this paper are the opinions of the authors and do not represent official positions of the Department of the Navy.

# Appendices

## A. Joint Feature Metric to combine different features using Error Analysis

The need of a metric to combine various features is important to reflect the performance of ensemble models where there are multiple features extracted through various techniques. We developed one such metric which looks at the error analysis of different features.

Let's consider a simple binary classification example where we have 100 test samples and let feature **A** mis-classify 10 of these samples, and feature **B** mis-classify 20 of the test samples. An ideal case for the features **A** and **B** to be jointly rewarding is when the overlap between the mis-classified samples is 0, i.e., when feature **B** can correctly classify the 10 samples that were mis-classified by feature **A**, and feature **A** can correctly classify the 20 samples that were mis-classified by feature **B**. We can then define an error-analysis matrix as follows: element $(i,j)$ represents the number of samples classified correctly by the feature representing $j^{th}$ column and mis-classified by the feature representing $i^{th}$ row. In our example, the error-analysis matrix (say $EA_I$) will then be:

$$EA_I = \begin{pmatrix} 0 & 10 \\ 20 & 0 \end{pmatrix}$$

Let us consider another case where the overlap between the mis-classified samples is 5, i.e., feature **B** correctly classifies 5 (out of 10) samples that were mis-classified by feature **A**, and feature **A** correctly classifies 15 (out of 20) samples that were mis-classified by feature **B**. In this case, the error-analysis matrix (say $EA_R$) will be:

$$EA_R = \begin{pmatrix} 0 & 5 \\ 15 & 0 \end{pmatrix}$$

The row-normalized (by number of mis-classified samples by the feature representing the row) matrices will then be:

$$\hat{EA}_I = \begin{pmatrix} 0 & 10/10 \\ 20/20 & 0 \end{pmatrix} = \begin{pmatrix} 0 & 1 \\ 1 & 0 \end{pmatrix}$$

$$\hat{EA}_R = \begin{pmatrix} 0 & 5/10 \\ 15/20 & 0 \end{pmatrix} = \begin{pmatrix} 0 & 0.5 \\ 0.75 & 0 \end{pmatrix}$$

The "closeness" between $\hat{EA}_I$ and $\hat{EA}_R$ is defined as the measure to quantify combining features **A** and **B**. This "closeness" is measured by a Matrix 2-norm, which is unique as the norm of the difference matrix can never exceed the value 2 (this is true for any square matrix). Therefore, this metric is well-defined as it belongs to a range [0,2]. A value close to 0 demonstrates the usefulness of the combination of features. A final confidence score (value between 0 and 1) called "Joint Feature Score ($JFS$)" which is the measure to quantify the combination of features can then be defined as follows:

$$JFS = \frac{2 - \|\hat{EA}_I - \hat{EA}_R\|_2}{2}$$

In our example,

$$JFS_{AB} = \frac{2 - \left\| \begin{pmatrix} 0 & 0.5 \\ 0.25 & 0 \end{pmatrix} \right\|_2}{2} = \frac{1.5}{2} = 0.75$$

which can be interpreted in two ways: (1) Features **A** and **B** can be combined to give better ensemble feature representation for classification with a confidence score of 0.75, (2) the maximum possible error reduction of the ensemble model (using features **A** and **B**) in terms of accuracy is 75% compared to the worse performing feature (**B**) model i.e., the ensemble model will have an estimated classification accuracy of 95% (80 + (0.75*20)) compared to 80% obtained using the feature **B** model.

This metric can also be computed for more than two features, where the error-analysis matrix will be a square matrix and the number of rows or columns are greater than two. In such cases, the joint feature measure between two or more features can be computed by extracting the corresponding rows/columns of the error-analysis matrix and then computing the $JFS$ value for those features.